\documentclass[11pt]{article}
\usepackage{epsf} 
\usepackage{amsmath}
\usepackage{amssymb}
\usepackage{epsfig}
\usepackage{latexsym}
\usepackage{amsfonts}
\usepackage{graphicx}%
\usepackage{varioref}
\usepackage{ifthen}
\begin{document}
\parindent 0mm 
\setlength{\parskip}{\baselineskip} 
\thispagestyle{empty}
\pagenumbering{arabic} 
\setcounter{page}{0}
\mbox{ }
\rightline{UCT-TP-279/2010}
\newline
\newline
\rightline{February  2010}
\newline

\noindent
\begin{center}
\Large \textbf{The Operator Product Expansion Beyond Perturbation Theory in QCD $^1$}
\end{center}
{\LARGE \footnote{{\LARGE {\footnotesize Invited talk at the XII Mexican Workshop on Particles \& Fields, Mazatlan, November 2009. To be published in American Institute of Physics Conference Proceedings Series.
Work supported in part by NRF (South Africa). }}}}

\begin{center}
C. A. Dominguez
\end{center}

\begin{center}
Centre for Theoretical Physics and Astrophysics\\
University of Cape Town, Rondebosch 7700, South Africa, and Department of Physics, Stellenbosch University, Stellenbosch 7600, South Africa
\end{center}

\begin{center}
\textbf{Abstract}
\end{center}

\noindent
The Operator Product Expansion (OPE) of current correlators at short distances beyond perturbation theory in QCD, together with Cauchy's theorem in the complex energy plane, are the pillars of the method of QCD sum rules. This technique provides an analytic tool to relate QCD with hadronic physics at low and intermediate energies. It has been in use for over thirty years to determine hadronic parameters, form factors, and QCD parameters such as the quark masses, and the running strong coupling at the scale of the $\tau$-lepton. QCD sum rules provide a powerful complement to numerical simulations of QCD on the lattice. In this talk  a short review of the method is presented for non experts, followed by three examples of recent applications.
\newpage
\section{Introduction}
The method of QCD sum rules, introduced by Shifman, Vainshtein, and Zakharov \cite{SVZ} more than thirty years ago, has become a powerful technique to study hadronic physics in the low energy resonance region by means of QCD \cite{REVIEW}. It is also a complementary tool to numerical simulations of QCD on a lattice. The range of applications has steadily grown over the years and covers the determination of the hadronic spectrum (masses, couplings, and widths), electromagnetic, weak, and strong form factors, quark masses and the strong coupling, the extension of QCD to finite temperature and its related topics of chiral symmetry restoration and quark-gluon deconfinement \cite{T}. It is also a tool to confront QCD predictions with experimental data, e.g. from $e^+ e^-$ annihilation and hadronic $\tau$-lepton decays. This method is based on two fundamental pillars: (i) the operator product expansion of current correlators at short distances, extended beyond perturbation theory to incorporate quark-gluon confinement, and (ii) Cauchy's theorem in the complex energy (squared) plane, often referred to as quark-hadron duality. To be more specific, let us consider a typical object in QCD in the form of the two-point function, or current correlator
\begin{equation}
\Pi(q^2)\,=\,i\; \int \,d^4 x \; e^{iqx} \; <0|\,T(J(x)\,J(0))\,|0 >,
\end{equation}
where the local current $J(x)$ is built from the quark and gluon fields entering the QCD Lagrangian, and it has definite quantum numbers. Equivalently, this current can be written in terms of hadronic fields having the same quantum numbers. The specific  choice of current will depend on the application one has in mind. For instance, if one is interested in determining the hadronic properties of the $\rho^+$-meson, then one would choose the QCD vector isovector current $J_\mu(x) = \bar{d}(x) \, \gamma_\mu \,u(x)$, and its hadronic realization in terms of the $\rho^+$-meson field. If the goal is to determine the values of the light quark masses, then the ideal object would be the correlator involving the axial-vector current divergences $J_5(x)|^i_j = (m_i + m_j) \bar{\psi}^i(x)  \gamma_5 \psi_j(x)$, with $i,j$ the up, down, or strange quark flavors. The hadronic representation of this correlator contains the pseudoscalar meson ($\pi$ or K) mass and coupling, followed by its radial excitations and the hadronic continuum. The tool to relate  these two representations is Cauchy's theorem in the complex energy (squared) plane, to be discussed shortly. \\
The QCD correlator, Eq.(1), will contain a perturbative piece (PQCD), computed up to a given loop order in perturbation theory, and a non perturbative part mostly reflecting quark-gluon confinement. The leading order in PQCD is shown in Fig.1.  Since QCD has never been solved analytically, the effects due to confinement can only be introduced by parameterizing quark and gluon propagator corrections effectively in terms of vacuum condensates. This is done as follows. In the case of the quark propagator
\begin{equation}
S_F (p) = \frac{i}{\not{p} - m}\;\;\Longrightarrow \;\;\frac{i}{\not{p} - m + \Sigma(p^2)} \;, 
\end{equation}
the quark propagator correction $\Sigma(p^2)$ would contain the information on confinement. One expects this correction to peak at and near the quark mass-shell, i.e. for $p \simeq 0$ in the case of light quarks. Effectively, this can be viewed as in Fig. 2, where the (infrared) quarks in the loop have zero momentum and interact strongly with the physical QCD vacuum. This effect is then parameterized in terms of the quark condensate $\langle 0| \bar{q}(0) q(0) | 0 \rangle$.
\begin{figure}[ht]
\begin{center}
  \includegraphics[height=.1\textheight]{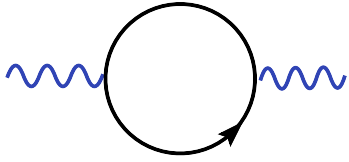}
  \caption{Leading order PQCD correlator. All values of the four-momentum of the quark in the loop are allowed. The blue wiggly line represents the current of momentum $q$ ($-q^2 >> 0$).}
  \label{fig:figure1}
  \end{center}
\end{figure}
Similarly, in the case of the gluon propagator one would have
\begin{equation}
D_F (k) = \frac{i}{k^2}\;\;\Longrightarrow \;\;\frac{i}{k^2 + \Lambda(k^2)} \;,
\end{equation}
where the gluon propagator correction will peak at $k\simeq 0$, and the effect of confinement in this case will be parameterized by the gluon condensate $\langle 0| \alpha_s\; \vec{G}^{\mu\nu} \,\cdot\, \vec{G}_{\mu\nu}|0\rangle$ (see Fig.2).
\begin{figure}[ht]
\begin{center}
\includegraphics[scale=0.8]{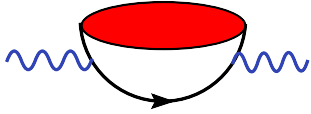}
\caption{Quark propagator modification due to (infrared) quarks interacting with the physical QCD vacuum, and involving the quark condensate. Large momentum flows through the bottom propagator.}
\label{fig:figure2}
\end{center}
\end{figure}
In addition to the quark and the gluon condensate there is a plethora of higher order condensates entering the OPE of the current correlator at short distances, i.e.
\begin{equation}
\Pi(q^2)|_{QCD}\,=\, C_0\,\hat{I} \,+\,\sum_{N=0}\;C_{2N+2}(q^2,\mu^2)\;\langle0|\hat{O}_{2N+2}(\mu^2)|0\rangle \;,
\end{equation}
where $\mu^2$ is the renormalization scale, and where the Wilson coefficients in this expansion depend on the Lorentz indices and quantum numbers of $J(x)$ and  of the local gauge invariant operators $\hat{O}_N$ built from the quark and gluon fields. These operators are ordered by increasing dimensionality and the Wilson coefficients, calculable in PQCD, fall off by corresponding powers of $-q^2$. 
Since there are no gauge invariant operators of dimension $d=2$ involving the quark and gluon fields in QCD, it is normally assumed that the OPE starts at dimension $d=4$ (with the quark condensate being multiplied by the quark mass). This is supported by results from QCD sum rule analyses of $\tau$-lepton decay data, which show no evidence of $d=2$ operators \cite{C2}.
\begin{figure}[ht]
\begin{center}
\includegraphics[scale=0.6]{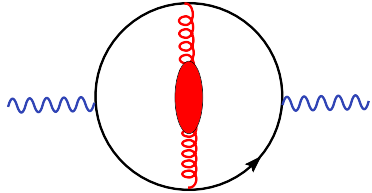}
\caption{Gluon propagator modification due to (infrared) gluons interacting with the physical QCD vacuum, and involving the gluon condensate. Large momentum flows through the quark propagators.}
\label{fig:figure3}
\end{center}
\end{figure}
The unit operator in Eq.(4) has dimension $d=0$ and $C_0 \hat{I}$ stands for the purely perturbative contribution. The Wilson coefficients as well as the vacuum condensates depend on the renormalization scale. In the case of the leading $d=4$ terms in Eq.(4) the $\mu^2$ dependence of the quark mass cancels the corresponding dependence of the quark condensate, so that this contribution is a renormalization group (RG) invariant. Similarly, the gluon condensate is also a RG invariant, hence once determined in some channel these condensates can be used throughout. At dimension $d=6$ there appears the four-quark condensate, obtained from Fig.1 at the next to leading order (one gluon exchange) and allowing all four quark lines to interact with the physical vacuum (see Fig.4). While this condensate has a residual renormalization scale dependence, this is so small that in practice it can be ignored. The four-quark condensate, while relatively small, is crucial to explain the large $\rho(770)$ - $a_1(1260)$ mass splitting. In most applications $- q^2$ is chosen large enough so that the condensates of higher dimension ($d \geq 8$) can be safely ignored.
\begin{figure}[ht]
\begin{center}
\includegraphics[scale=0.6]{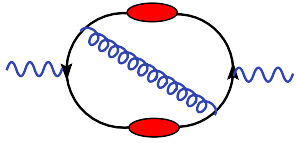}
\caption{The four-quark condensate of dimension $d=6$ in the OPE. This is responsible for the $\rho - a_1$ mass splitting. Large momentum flows through the gluon propagator.}
\label{fig:figure4}
\end{center}
\end{figure}
The numerical values of the vacuum condensates cannot be calculated analytically from first principles as this would be tantamount to solving QCD exactly.
One exception is that of the quark condensate which enters in the Gell-Mann-Oakes-Renner relation, a QCD low energy theorem following from the global chiral symmetry of the QCD Lagrangian. Otherwise, it is possible to extract values for the leading vacuum condensates using QCD sum rules together with experimental data, e.g. $e^+ e^-$ annihilation into hadrons, and hadronic decays of the $\tau$-lepton. Alternatively, as lattice QCD  improves in accuracy it should become a valuable source of information on these condensates.\\
Turning to the hadronic sector, bound states and resonances appear in the complex energy (squared) plane (s-plane) as poles on the real axis, and singularities in the second Riemann sheet. In addition there will be multiple cuts reflecting non-resonant multi-particle production. All these singularities lead to a discontinuity across the positive real  axis. Choosing an integration contour as shown in Fig.5, and given that there are no further singularities in the complex s-plane, Cauchy's theorem leads to the finite energy sum rule (FESR)
\begin{figure}[ht]
\begin{center}
  \includegraphics[height=.25\textheight]{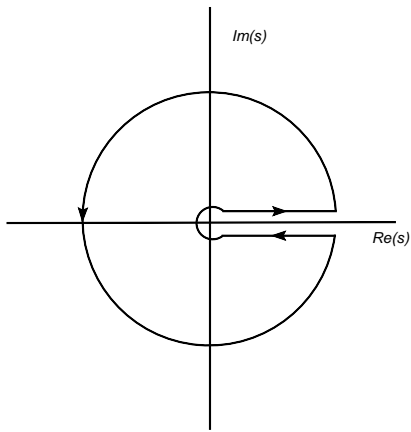}
  \caption{Integration contour in the complex s-plane. The discontinuity across the real axis brings in the hadronic spectral function, while integration around the circle involves the QCD correlator.}
\label{fig:figure5}
\end{center}
\end{figure}
\begin{equation}
\int_{\mathrm{sth}}^{s_0} ds\; \frac{1}{\pi}\; f(s) \;Im \Pi(s)|_{HAD} \; = \; -\, \frac{1}{2 \pi i} \; \oint_{C(|s_0|) }\, ds \;f(s) \;\Pi(s)|_{QCD} \;,
\end{equation}
where $f(s)$ is an arbitrary (analytic) function, $s_{th}$ is the hadronic threshold, and the finite radius of the circle, $s_0$, is large enough for QCD and the OPE to be used on the circle. Physical observables determined from FESR should not depend on $s_0$. In practice, though, 
this independence is not exact, and there is usually a region of stability where
observables are fairly independent of $s_0$, typically in the range $s_0 \simeq 1 - 4 \; \mbox{GeV}^2$. The variation of an observable in the stability region is incorporated into the error of the determination.
Equation (5) is the mathematical statement of what is usually referred to as quark-hadron duality. Since QCD is not valid in the time-like region ($s \geq 0$), in principle there is a possibility of problems on the circle near the real axis (duality violations). I shall come back to this issue later. The right hand side  of this FESR involves the QCD correlator which is expressed in terms of the OPE as in Eq.(4). The left hand side calls for the hadronic spectral function which is written as
\begin{equation} 
Im \Pi(s)|_{HAD}\,=\, Im \Pi(s)|_{POLE}\,+\, Im \Pi(s)|_{RES} \,\theta(s_0-s)\,+\, Im \Pi(s)|_{PQCD}\,\theta(s-s_0) \;,
\end{equation}
where the ground state pole, absent in some channels, is followed by the resonances which merge smoothly into the hadronic continuum above some threshold $s_0$. This continuum is expected to be well represented by PQCD if $s_0$ is large enough. Due to this, if one were to consider an integration contour in Eq.(5) extending to infinity, the cancellation between the hadronic continuum in the left hand side and the PQCD contribution to the right hand side, would render the sum rule a FESR. The performance of the contour integral in the complex s-plane is discussed in the next section.
\section{Finite energy sum rules}
The integration in the complex s-plane of the QCD correlator is usually carried out in two different ways, Fixed Order Perturbation Theory (FOPT) and Contour Improved Perturbation Theory (CIPT) \cite{CIPT}. The first method treats running quark masses and the strong coupling as fixed at a given value of $s_0$. After integrating all logarithmic terms ($\ln(-s/\mu^2)$) the RG improvement is achieved by setting the renormalization scale to $\mu^2 = - s_0$. In CIPT the RG improvement is performed before integration, thus eliminating logarithmic terms, and the running quark masses and strong coupling are integrated (numerically) around the circle. This requires solving numerically the RGE for the quark masses and the coupling at each point on the circle.
The FESR Eq.(5), with $f(s)=1$,  in FOPT can be written as
\begin{equation}
(-)^N \, C_{2N+2} \, \langle0| \hat{O}_{2N+2}|0 \rangle =  \int_0^{s_0} \,ds\, s^N \, \frac{1}{\pi}\, Im \,\Pi(s)|_{HAD} \,-\, s_0^{N+1} \; M_{2N+2}(s_0) \, ,
\end{equation}
where the dimensionless PQCD moments $M_{2N+2}(s_0)$ are given by
\begin{equation}
M_{2N+2}(s_0) = \frac{1}{s_0^{(N+1)}} \, \int_0^{s_0}\, ds\,s^N \, \frac{1}{\pi} \, Im \, \Pi(s)|_{PQCD}\;.
\end{equation} 
If the hadronic spectral function is known in some channel from experiment, then $Im \Pi(s)|_{HAD} \equiv Im \Pi(s)|_{DATA}$, and Eq.(7) can be used to determine the values of the vacuum condensates. Subsequently, Eq.(7) can be used in a different channel to determine the masses and couplings of the hadrons in that channel. It is important to mention that the correlator $\Pi(q^2)$ is generally not a physical observable. However, this has no effect in FOPT as the unphysical constants in the correlator do not contribute to the integrals. The situation is quite different in CIPT where Eq.(5) cannot be used for unphysical correlators. For instance for a correlator whose physical counterpart is the second derivative (needed to eliminate a first degree polynomial),  Cauchy's theorem and the resulting FESR must be written for the second derivative. In this case  one has to use the following identity \cite{DNRS1}
\begin{equation}
\oint_{C(|s_0|) }\, ds \, g(s) \, \Pi(s) = \oint_{C(|s_0|) } \, ds \, [F(s) - F(s_0)] \;\Pi''(s) \;,
\end{equation}
where
\begin{equation}
F(s) = \int ^s ds' \left[ \int^{s'} ds'' g(s'') - \int^{s_0} ds'' g(s'')\right] \;,
\end{equation}
and $g(s)$ is an arbitrary function. This is easily proved by integrating by parts the right hand side of Eq.(9) and using Eq.(10) to obtain the left hand side. In this case Eq.(5) becomes
\begin{equation}
\int_{\mathrm{sth}}^{s_0} ds\; g(s)\;\frac{1}{\pi}\; Im \Pi(s)|_{HAD} \; = \; -\, \frac{1}{2 \pi i} \; \oint_{C(|s_0|) }\, ds \; [F(s)-F(s_0)] \;\Pi''(s)|_{QCD} \;,
\end{equation}
which is the master FESR to use in CIPT. The running quark masses and the running strong coupling entering $\Pi''(s)$  are now functions of the integration variable and are not fixed as previously in FOPT. The running coupling obeys the RGE
\begin{equation}
s \; \frac{d \, a_s(-s)}{d s} = \beta (a_s) = - \sum_{N=0} \beta_N \; a_s(-s)^{N+2} \;,
\end{equation}
where $a_s \equiv \alpha_s/\pi$, and  e.g. for three quark flavors $\beta_0 = 9/4$, $\beta_1 = 4$, $\beta_2 = 3863/384$,
$\beta_3 = (421797/54 + 3560 \zeta(3))/256$. In the complex s-plane $s = s_0\, e^{ix}$ with the angle $x$ defined in the interval $x \in (- \pi, \pi)$. The RGE then becomes
\begin{equation}
\frac{d \, a_s(x)}{d x} = - i \sum_{N=0} \beta_N \; a_s(x)^{N+2} \;,
\end{equation}
This RGE can be solved numerically at each point on the integration contour of Eq.(11) using e.g. a modified Euler method, providing as input $ a_s (x=0) = a_s (- s_0)$. Next, the RGE for the quark mass is given by
\begin{equation}
\frac{s}{m} \; \frac{d \, m(-s)}{d s} = \gamma (a_s) = - \sum_{M=0} \gamma_M \; a_s^{M+1} \;,
\end{equation}
where e.g. for three quark flavors $\gamma_0 = 1$, $\gamma_1 = 182/48$, $\gamma_2 = [8885/9 - 160 \,\zeta(3)]/64$, $\gamma_3 = [2977517/162 - 148720 \,\zeta(3)/27 + 2160 \,\zeta(4) - 8000\, \zeta(5)/3]/256$. With the aid of Eqs. (12)-(13) the above equation can be converted into a differential equation for $m(x)$ and integrated, with the result
\begin{equation}
m(x) = m(0) \;exp \Big\{ - i \int_0^x dx' \sum_{M=0} \gamma_M \, [a_s(x')]^{M+1}\Big\}\;,
\end{equation}
where the integration constant $m(0)$ can be  identified with $m(s_0)$.
\section{APPLICATIONS}
As a first application I sketch the determination of the strong coupling at the scale of the $\tau$ mass \cite{PICH} using experimental data on the ratio $R_\tau = \Gamma(\tau \rightarrow hadrons)/\Gamma(\tau \rightarrow leptons)$. The $\tau$-decays proceed through the $V-A$ current, so that  $J(x)$ in Eq.(1) becomes $J(x) \rightarrow J_\mu(x) = \bar{d}(x) \gamma_\mu (1 - \gamma_5) u(x)$ in the non-strange sector. The ratio $R_\tau$ can then be written as
\begin{equation}
R_\tau = constant \, \int_0^{M_\tau^2}\,\frac{ds}{s_0} \, (1 - \frac{s}{s_0})^2
\left[ (1 + 2 \frac{s}{s_0})\, Im \Pi^{(1)}(s) + Im \Pi^{(0)}(s)\right] \;,
\end{equation}
where $\Pi^{(J)}$ stands for the spin $J=0$ or $J=1$ components. The QCD spectral functions are known in PQCD up to five-loop order, and the non-perturbative contribution is very tiny as it only starts at dimension $d=6$ (due to the presence of the  integration kernel). Hence, the strong coupling is basically the only unknown which can be determined after using the experimental data for the left hand side of the FESR above. The latest results are $\alpha_s(M_\tau^2) = 0.332 \pm 0.016$ from \cite{BAIKOV} using FOPT and CIPT, and $\alpha_s(M_\tau^2) = 0.344 \pm 0.009$ from \cite{ALEPH} using CIPT. Computing the evolution of the coupling to higher energies and comparing with independent determinations at the scale of the $Z$ mass provides the most accurate test of asymptotic freedom. It also provides the most accurate and transparent  determination of $\alpha_s$ at the five-loop level.\\
Next I discuss an application where the integration kernel $f(s)$ in Eq.(5) is of great importance \cite{FPI}. In the axial-vector channel, the FESR Eq.(7) with $f(s)=1$ and $N=0$ can be confronted with data from $\tau$-decay. The hadronic
\begin{figure}[ht]
\begin{center}
  \includegraphics[height=.28\textheight]{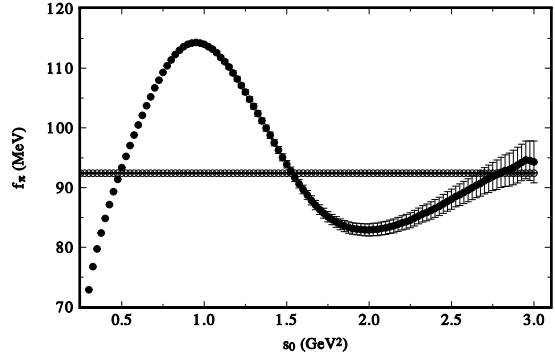}
  \caption{Results for $f_\pi$ from the standard  FESR in the axial-vector channel, Eq. (7) with N=0, with no dimension $d=2$ term, and using  CIPT, with $\Lambda = 365 \,\mbox{MeV}$ ($\alpha_s(M_\tau^2)= 0.335$). The straight line  is the experimental value of $f_\pi$, and the points are the integrated data with the experimental errors.}
  \label{fig:figure6}
  \end{center}
\end{figure}
spectral function is then written as the sum of the pion pole and the resonance data known up to the kinematical end point $s_0 = M_\tau^2$. The moment $M_2(s_0)$ is known up to five-loop order in PQCD, so that the FESR can be used to confront the resonance data plus PQCD with e.g. $f_\pi$. As seen from Fig. 6 the agreement is rather poor, except possibly near the end point. At first sight, this may be interpreted as a signal for quark-hadron duality violations near the real axis even at high enough energy. In fact, it has been known for quite some time that e.g. the Weinberg (chiral) sum rules are not saturated by the $\tau$ decay data unless one introduces {\it pinched} integration kernels, e.g.  $f(s) = [1 - (s/s_0)]^{(N+1)}$ \cite{PINCH}. Unfortunately, the $\tau$-lepton is not massive enough to probe higher energy regions. In spite of this it is still possible to enlarge the energy range by introducing as integration kernel a polynomial $f(s) \equiv P(s)$ designed to eliminate the (unknown) hadronic contribution to the integral between $s_1$ and $s_0 \geq s_1$, where $s_1$ is at or near the end point of the data. It has been shown \cite{FPI} that the optimal degree of $P(s)$ is the simplest, i.e. the linear function
\begin{equation}
P(s)=1-\frac{2s}{s_{0}+s_{1}} \,,
\end{equation}
so that
\begin{equation}
 \mbox{constant} \times \int_{s_1}^{s_0} P(s) ds  = 0\,.
\end{equation}
In this case the complete FESR becomes a linear combination of a dimension-two and a dimension-four FESR, which from Eqs.(7) and (17) it is given by
\begin{eqnarray}
2 \, f_\pi^2 &=& - \int_{0}^{s_{1}} ds \, P(s) \, \frac{1}{\pi}\, Im \Pi(s)|_{DATA}
+ \frac{s_0}{4 \pi^2} \left[ M_2(s_0) - \frac{2 s_0}{s_0+s_1} M_4 (s_0) \right] \nonumber \\[.3cm]
&+& \frac{1}{4 \pi^2} \left[ C_2 \langle \hat{O}_2 \rangle +\frac{2}{s_0+s_1} C_4 \langle \hat{O}_4 \rangle \right] \, + \Delta(s_0)\,,
\end{eqnarray}
where the pion pole has been separated from the data, and the chiral limit is understood. The term $\Delta(s_0)$ is the error being made by assuming that the data is constant in the interval $s_1 - s_0$. It is possible to estimate this error which turns out to be two to three orders of magnitude smaller than  $2 f_\pi^2$ on the left hand side of Eq.(19) \cite{FPI}.
\begin{figure}[ht]
\begin{center}
  \includegraphics[height=.3\textheight]{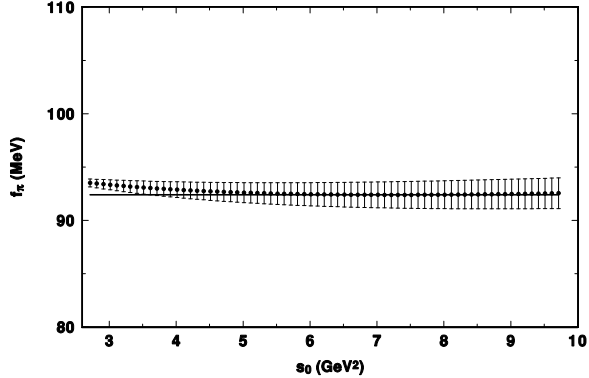}
  \caption{Results for $f_\pi$ from the FESR in the axial-vector channel, Eq. (19), with $C_2\langle{\hat{O}}_2\rangle =0$, $C_4\langle{\hat{O}}_4\rangle = 0.05 \,\mbox{GeV}^4$, , and using  CIPT  with $\Lambda = 365 \,\mbox{MeV}$ ($\alpha_s(M_\tau^2)= 0.335$). The straight line  is the experimental value of $f_\pi$, and the points with errors are the integrated data up to $s_1\simeq M_\tau^2$.}
  \label{fig:figure7}
\end{center}  
\end{figure}
As can be seen from Fig. 7 the FESR Eq.(19) shows an excellent consistency between QCD and the $\tau$ data in the axial-vector channel in a remarkably wide region $s_0 \simeq 4 \, -\,10\, \mbox{GeV}^2$. A similar consistency is also found in the vector channel, where QCD is now confronted with zero (there is no pole in this channel). This result shows no evidence for quark-hadron duality violations in the vector and axial-vector channels. In addition, this method can prove quite useful in QCD FESR applications needing an extension of the energy region of available data.\\ 
Finally, I discuss the use of FESR to determine the values of the QCD light quark masses \cite{DNRS1}. As mentioned in the Introduction, in the case of the  up-,down, and strange-quark masses the ideal current operator in Eq.(1) is the axial-vector current divergence. Its advantage is that the masses appear here as overall multiplicative factors, rather than as subleading power corrections like in other correlators, e.g. the vector or axial-vector correlators. The great disadvantage is that there is no direct experimental data beyond the pseudoscalar meson poles, i.e. the hadronic resonance spectral function, $Im \,\Pi(s)|_{RES}$ in Eq.(6) is not known experimentally. The only available information  is that there are two radial excitations in the non-strange ($\pi$) as well as in the strange ($K$) channel with known masses and widths. This is hardly enough to reconstruct the full spectral function. In fact, inelasticity, non-resonant background, and resonance interference are impossible to guess so that a model is needed for the resonant spectral function. This fact, which introduces a serious systematic uncertainty, has affected all quark mass determinations using QCD sum rules until recently \cite{DNRS1}. The breakthrough has been to introduce an integration kernel in the FESR tuned to suppress substantially the resonance energy region above the ground state. This kernel is of the form $f(s)\equiv \Delta_5(s) = 1\, - \,a_0\,s \,-\,a_1\,s^2$, where the coefficients are fixed by requiring that $\Delta_5(s)$ vanish at the peak of the two radial excitations, i.e. $\Delta_5(M_1^2) = \Delta_5(M_2^2) = 0$. This has the effect of reducing the resonance contribution to the FESR to a couple of a percent of the ground state contribution, well below the uncertainty due to $\alpha_s$. Hence the systematic uncertainty is essentially eliminated and one finds, using FOPT and CIPT 
\begin{eqnarray}
m_u (2 \;\mbox{GeV}) = 2.9 \; \pm \; 0.2 \; \mbox{MeV} \;, \nonumber \\ [.1cm]
m_d (2 \;\mbox{GeV}) = 5.3 \; \pm \; 0.4 \; \mbox{MeV} \;, \nonumber \\ [.1cm]
m_{ud} \equiv \frac{m_u + m_d}{2} = 4.1 \; \pm \; 0.2 \; \mbox{MeV}\;, \nonumber\\ [.1cm]
m_s (2 \;\mbox{GeV}) = 102 \; \pm \; 8\; \mbox{MeV} \;. \nonumber 
\end{eqnarray}
This is at present the most accurate determination of the light quark masses using QCD sum rules.


\end{document}